# 5G enabled Mobile Edge Computing security for Autonomous Vehicles

Daryll Ralph D'Costa, Dr. Robert Abbas, **Macquarie University**

daryllralph.dcosta@students.mq.edu.au, robert.abbas@mq.edu.au

## ABSTRACT

The world is moving into a new era with the deployment of 5G communication infrastructure. Many new developments are deployed centred around this technology. One such advancement is 5G Vehicle to Everything communication. This technology can be used for applications such as driverless delivery of goods, immediate response to emergencies and improving traffic efficiency. The concept of Intelligent Transport Systems (ITS) is built around this system which is completely autonomous. This paper studies the Distributed Denial of Service (DDoS) attack carried out over a 5G network and analyses security threats. The aim is to implement a machine learning model capable of classifying different types of DDoS attacks and predicting the quality of 5G latency.

The initial steps of implementation involved the synthetic addition of 5G parameters into the dataset. Subsequently, the data was label encoded, and minority classes were oversampled to match the other classes. Finally, the data was split as training and testing, and machine learning models were applied. Although the paper resulted in a model that predicted DDoS attacks, the dataset acquired significantly lacked 5G related information. Furthermore, the 5G classification model needed more modification. The research was based on largely quantitative research methods in a simulated environment. Hence, the biggest limitation of this research has been the lack of resources for data collection and sole reliance on online data sets. Ideally, a Vehicle to Everything (V2X) project would greatly benefit from an autonomous 5G enabled vehicle connected to a mobile edge cloud. However, this project was conducted solely online on a single PC which further limits the outcomes. Although the model underperformed, this paper can be used as a framework for future research in Intelligent Transport System development.

## INTRODUCTION

The world is advancing into a new era of autonomous self-driving vehicles. A lot of research is currently underway to improve the functionality of these vehicles and make them safe. With the emergence of 5G cellular networks, the autonomous self-driving industry can leverage this infrastructure to improve autonomous driving technology.

Human mistake is responsible for over 94 percent of all car accidents whereas mechanical failure is responsible for only 2% of all accidents [1]. While taking a closer look, it is found that speeding is responsible for one out of every four deaths [1]. The figures are stunning, revealing that a quarter of all traffic fatalities and most traffic accidents are avoidable. As a result, connected vehicles will be a revolutionary in terms of traffic safety.

Engineers are trying to implement an Intelligent Transport System (ITS) where vehicles communicate with each other and their surroundings completely on their own. This requires the vehicles to be equipped with high grade sensors and reliable wireless technology.

Autonomous vehicles are usually equipped with sensors like Light Detection and Ranging (LiDAR), Global Positioning System (GPS), radar and odometry that collect data and send it to the onboard processing unit. Additionally, the vehicles are also equipped with high-resolution cameras record everything. Subsequently, these vehicles then use image processing and computer vision to understand their surroundings and make decisions using machine learning. If the automobile cannot recognize a specific movement or pattern, it can send the data to the cloud. The cloud is a collection of vast to nearly unlimited resources for computing and storage.

5G networks have a component called mm Waves which enable Ultra-Reliable Low Latency Communication (URLLC) which allows signaling speeds of less than one millisecond. Additionally, cellular networks employ the concept of network slicing that reserves a dedicated frequency spectrum for Vehicle to Everything (V2X) based communication.

Malicious attackers can use these low latency mm Wave frequencies for wrong reasons. Security is the most critical aspect in self driving since human life is involved. Attackers try to manipulate, modify, corrupt, or block communication to the intended vehicle which can cause mishaps to happen on the road. One such attack is Distributed Denial of Service (DDoS) which aims to overload the receiving interface at the car and render it blind. This could potentially disrupt a specific vehicle or a group of vehicles and cause accidents. A possible solution is to employ machine learning principles to detect an attack and mitigate it.

# BACKGROUND AND RELATED WORK

This section explores the different aspects of the 5G V2X environment. It explains how vehicles communicate within themselves and to the infrastructure and environment. It also explains the importance of mobile edge cloud and the enabling aspects. Since this research discusses vehicular 5G security, it studies all types of attacks. Since the attack scenario is DDoS attacks, the thesis explains the Distributed Denial of Service attacks and their types. In addition, it discusses a brief overview of the various machine learning models. Finally, the thesis explains information on the dataset and other details of its implementation.

## 1.    Mobile Edge Computing

Cloud computing is an IT paradigm that allows users global access to shared pools of configurable system resources and higher-level services deployed rapidly and with less administrative effort over the Internet. Cloud computing, similar to a public resource, relies on shared resources to achieve continuity and profits.

MEC is a group of edge servers that are a type of edge device that serves as a network entrance point. Additionally, routers and network switches are examples of other edge devices. Edge devices frequently positioned within Internet exchange points (IxPs) allow multiple networks to join and share bandwidth [2].

The final aspect of MEC is customer premises equipment (CPE). The CPE is a part of the CSP's wide-area network (WAN) infrastructure provided to its enterprise users. It is located at a client location and ends WAN links into that location. Its goal is to provide stable and secure connectivity between WAN connected devices and the local area network (LAN) at that specific location. As a result, a CPE usually includes switching and smart routing technology for navigating traffic between the WAN and the LAN. In addition, these smart devices can perform load balancing between WAN links when multiple links are present. The security system installed provides industry-standard security services to traffic moving into and out of the organization. The switching, routing and firewall functions are usually a group of policies that offer Quality of Service (QoS) distinction and integration with enterprise policy systems and other features [3].

Many cloud-based companies and vendors take part in the Mobile Edge Computing ecosystem. Google, Amazon Web Services (AWS) and Microsoft Azure are examples of cloud companies that provide the necessary resources and services. They are Hyperscale Cloud Providers (HCP's) and actively compete to take over the edge computing market.

## 2.    Enabling Technologies

In recent years various technologies devised have further improved existing hardware. These technologies enable even faster communication of vehicular networks. In addition, they are abstracted on top of existing hardware and use their resources and work as individual software instances. It is a powerful tool, as it allows multiple separate instances of the software to run completely different tasks without being dependent on additional resources. Consequently, the hardware configured allocates a set amount of each technology. Some of the technologies mentioned are listed below.

### 2.1. Network Function Virtualization (NFV)

By employing full-blown virtualization technology, NFV improves how network operators construct their infrastructure. It is carried out by decoupling software instances from hardware platforms and decoupling functionality from the location for faster networking service provisions. In other words, NFV virtualizes network functions and runs them on existing hardware using software virtualization techniques (i.e., industry servers, switches, and storage units. Furthermore, virtual appliances launched on-demand do not require the need for additional hardware [4].

While in use, NFV technology in 5G V2X networks provides the availability of the IOS that makes it simple and easy to update the system by simply updating the software. The feasibility and viability of Vehicle-to-vehicle (V2V) collaboration in 5G V2X networks is due to Virtual Network Functions (VNF) reuse and migration. Additionally, VNF reuse enables efficient cluster-based networking. Communication delay is reduced by allowing adjacent vehicles to act as local VNF managers instead of faraway service centers [5].

### 2.2. Software Defined Networks

SDN (software-defined networking) is a new network concept that combines logic, centralised control, and programmability in one package. The SDN controller used to manage network resources flexibly centralises network data and integrates diverse network protocols and standards [6].

### 2.3. Smart Collaborative Networking

To create smart Internet architectures, Smart Collaborative Networking (SCN) was introduced. SCN believes that the future Internet narrative should switch from the existing service provisioning through controlled ownership of infrastructures to a future of unified management systems full of dynamic control and coordination between users, networks, and facilities through virtualization technology and programmability [7].

### 2.4. Blockchain

In 2008, blockchain technology was developed to enable the creation of Bitcoin. It's essentially a collection of rules that govern how the virtual currency works. The cryptocurrency was developed and traded further, and the community that developed the technology began to work on fixing bugs in the system. It is a communal registration system that consists of a chain of blocks, thus the name. This implies that rather than being kept in a single location, all blockchain data is spread across the numerous computing units linked to it [8].

## 3. DDoS attack types

There are many forms of DDoS attacks. Each attack type aims to exploit a particular weakness in the networking protocols. The overall goal of DDoS attacks is to overload networking equipment with traffic causing congestion and disabling the correct entity from accessing information. This attack can be even more dangerous when coupled with 5G. The super-fast speed of 5G networks will make it even harder for network equipment to handle DDoS traffic.

### 3.1. NTP based attack

An attacker exploits Network Time Protocol (NTP) server functionality to flood a specific client-server or other networks with an increasing amount of User Datagram Protocol (UDP) data traffic in this DDoS attack feedback. This attack has the potential to render the destination and its network infrastructure inaccessible to regular traffic [9].

### 3.2. DNS based attack

An attacker uses a Botnet to create a high number of resolution requests to a designated Internet Protocol (IP) address in this DDoS assault based on reflection [9]. Modern Domain Name System (DNS) applications rely on the DNS Security Extensions' sophisticated security protocols. Vehicular DNS communications require validation of the source and legitimacy of DNS resource entries. A solution to this attack is employing multi hop Vehicle-to-Vehicle communications to reach a name server, as well as a new Bloom filter-based technique to record verification [10].

### 3.3.    LDAP based attack

It is a reflection-based DDoS attack. Here the attacker sends requests to a publicly accessible insecure Lightweight Directory Access Protocol (LDAP) server to create amplified answers mirrored to a target server [9].

### 3.4.    MSSQL based attack

It is a reflection-based DDoS attack. Here the attacker uses the Microsoft SQL Server Resolution Protocol (MC-SQLR) to execute scripted requests with a faked IP address to appear as it was coming from the target server [9].

### 3.5.    NetBIOS based attack

It is a reflection-based DDoS attack. Here the attacker delivers a victim machine fraudulent "Name Release" or "Name Conflict" notifications to block all Network Basic Input/Output System (NetBIOS) network traffic [9].

### 3.6.    SNMP based attack

It is a massive DDoS attack that leverages the Simple Network Management Protocol (SNMP) to produce attack volumes of thousands of gigabits per second to clog the target's network pipes [9]. SNMP is used to configure and gather information from network devices such as servers, switches, routers, and printers. Additionally, it applies to SNMP reflected amplification attacks. The attacker uses SNMP to provide a flood of answers to the victim, like other reflection attacks. The offender sends many SNMP queries with a fabricated IP address (the target's) to numerous connected devices, which respond to the fraudulent address [11].

### 3.7.    SSDP based attack

Simple Service Discovery Protocol (SSDP) is a reflection-based DDoS attack. Here the attacker uses Universal Plug and Play (UPnP) networking protocols to transmit an amplified amount of traffic to a targeted victim [9].

### 3.8.    UDP Lag based attack

The goal of this attack is to use IP packets containing UDP datagrams to slow down or interrupt the targeted host [9].

### 3.9.    Web DDoS based attack

The goal of this attack is to hack into a Web server or application using malware takes advantage of valid HyperText Transfer Protocol (HTTP) GET or POST queries [9].

### 3.10.    SYN based attack

This attack makes use of the standard Transmission Control Protocol (TCP) three-way handshake (i.e., transmitting SYN (synchronize), SYN-ACK (synchronize-acknowledge), and responding with an ACK (acknowledge)) to consume resources and render the targeted network server unavailable [9].

### 3.11.    TFTP based attack

This attack uses TFTP servers connected to the internet to exploit the Trivial File Transfer Protocol (TFTP). Specifically, an attacker sends a file request to the victim TFTP server, which returns the data to the requesting target host [9].

### 3.12.    Port Scan based attack

This attack conducts a network security audit by scanning ports on a single machine or a network. The Queries in use detect which services are running on a remote server during the scanning process [9].

# DATA ANALYSIS AND VISUALIZATION

The dataset was downloaded from [12]. The dataset consisted of a ZIP file which in turn consisted of twelve folders. Each folder was a simulation of a specific kind of DDoS attack. After unpacking the data, the first step involved loading the data into a python Jupyter notebook. Since each dataset was too large, a total of 2200000 samples from each dataset were imported to not overload the operating PC environment. The twelve unique datasets were then merged to form a single dataset that contained all the DDoS attacks. On initial inspection of the dataset, it is observed that there was a total of 88 columns representing a distinct feature. The resultant dataset consisted of 5500000 unique data points. The dataset consisted of a Label column which described the type of DDoS attack. The initial unique Label counts are as follows:

```
DrDoS_SSDP       49989
Syn              49983
DrDoS_SNMP       49975
TFTP             49970
DrDoS_NetBIOS    49969
DrDoS_UDP        49964
DrDoS_MSSQL      49964
DrDoS_LDAP       49961
DrDoS_DNS        49958
UDP-lag          49454
DrDoS_NTP        49409
BENIGN            1350
WebDDoS             54
```

The dataset reveals a good distribution of most forms of attacks. The only attack which has a smaller number of recorded observations is Web DDoS attacks. Also, since most attacks are focused around simulating the attack, there are also very few recorded observations of Benign data. Consequently, the lack of these observations is handled in data pre-processing stage and the number of recorded observations is synthetically increased.

The dataset is a pure simulation of only DDoS attacks only. This means that there are no features corresponding to 5G. Since this thesis focuses on the aspects of 5G networks, and the goal is to simulate a machine learning base in a 5G Mobile Edge Computed environment, the dataset was updated to include information of 5G based parameters as well. This was done by making three new feature columns namely, 5G_RSRP, 5G_RSRQ and 5G_Latency.

## 1. Data visualizations

To better understand the distribution of the DDoS attack types, a pie chart is created to see the percentage of each attack. Figure 1 is the resultant pie chart showing the necessary information with respect to DDoS attacks. The analysis shows that most of the DDoS attacks are nicely balanced. Each once consists of an average of 9 percent of unique data. Except of Web DDoS attack which consists of 0.01 percent. Web DDoS attacks are generally based off the application layer [13], so it is common for the number of samples to be lesser. The pie chart analysis also shows that there is less benign data since most of the unique samples are simulated to be a part of DDoS attack. Of all the data, the NetBIOS, NTP, TFTP and Syn attacks have the highest distribution of samples.

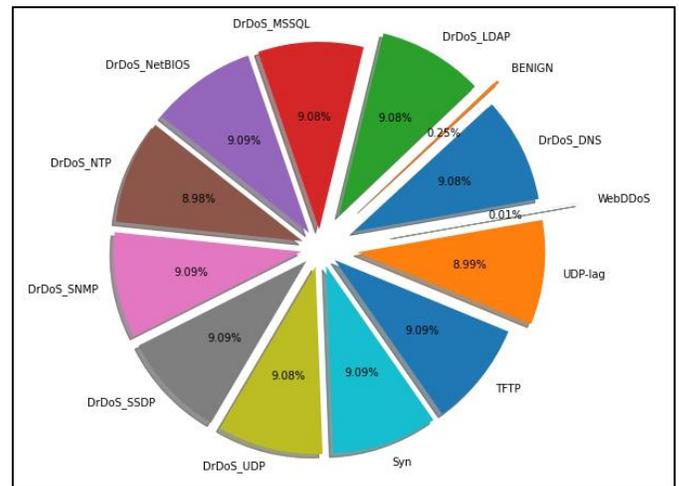

*Figure 1: Pie chart of DDoS attacks*

Since there is a total of over eighty features, only a few of them are explained. The explained features are in accordance with the K best features selected.

### 1.1. Source Port

The source port is the point of origin of an IP address. It is uniform across the TCP/IP and UDP protocol stacks. Figure 2 shows the distribution of source ports and the number ports being used. Analysis of the plot shows that the common ports for application layer access are used in benign. Generally, port 80 for HTTP access is used. It is observed that the other attacks use port numbers starting from roughly 5000 and end up using almost all the ports till over 60,000. TFTP in particular has higher counts of port usage.

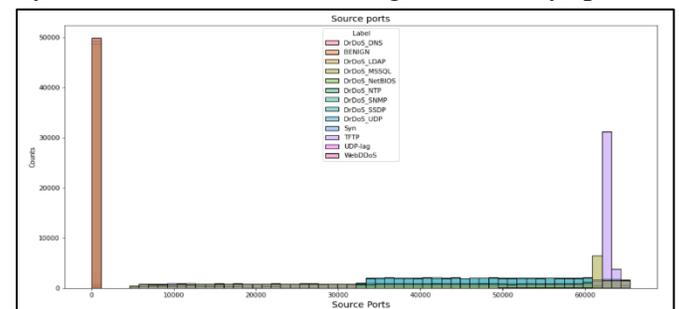

*Figure 2: Source ports*

### 1.2. Destination Port

Similar to source ports, the destination port is the final destination the IP packed is supposed to reach. This is also uniform cross the TCP/IP and UDP protocols. Figure 3 shows the distribution of destination ports and the number of them being used. From the plot, it is evident that almost all the ports are being occupied by the various attack types. The count for each port is also very high. This means that from one source port, the attack type tries to attack multiple destination port as evident in the plot. Only benign data is being transported over the pre-defined application layer ports.

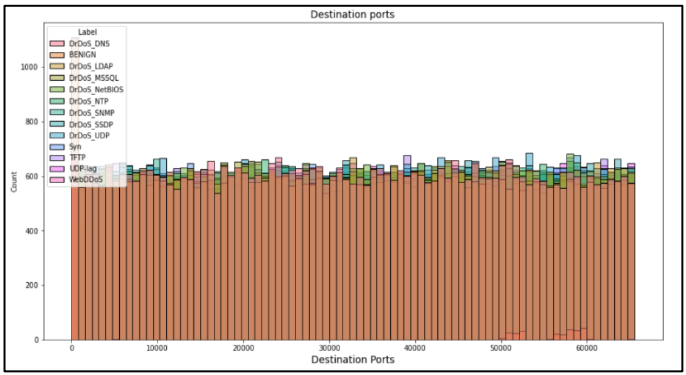

Figure 3: Destination ports

### 1.3. 5G Latency Label

The 5G latency was synthetically added. The latency was further classified as good or bad with respect to attack type. The label good is given to latencies if the latency value is below 30 ms. All other latencies are classified as bad as the DDoS attacks may be affecting the 5G radio system. Figure 4 shows the counts of good 5G latency vs bad 5G latency. This 5G latency label is one of the output labels that will predicted using machine learning. From the counts we can see that most of the counts come under the bad category. Machine learning models will become biased if the current dataset was to be passed. The minority is balanced out in a later stage to give the models a fair chance to analyze the data.

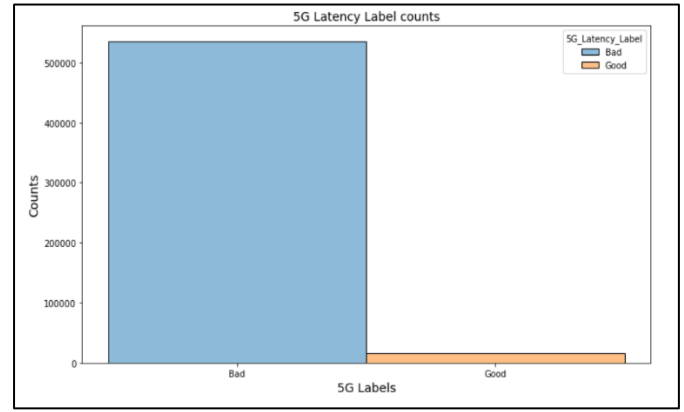

Figure 4: 5G Latency Label (Good vs Bad)

## DATA PRE-PROCESSING

### 1. Label encoding

Machine learning models struggle to process anything other than numbers. The goal is to predict the DDoS attack and 5G Labels but is not possible if they are of categorical form. Until this stage, each DDoS attack has a name associated such as DDoS_NTP. Label encoding converts these labels into a unique number. The process of label encoding is applied to both the DDoS labels as well as to 5G Latency Labels. The figure 10 shows the code used to convert the labels into numerical labels.

### 2. Handling categorical, infinite, and unwanted data

Once the encoding process is done the column with categorical labels for both DDoS and 5G Latency values must be dropped. The data set also consists of infinite values which need to be handled properly. A small function is written which takes in the data frame columns as values processes at and returns the columns which have infinite values. There are also some categorical labels that need to be dropped. Additionally, there are some other features like unnamed:0, source port and destination port which are added do a final list of columns that need to be dropped. Consequently, all the unwanted columns are grouped together and dropped using the pandas drop method.

### 3. Performing oversampling to rectify minority classes

As discussed in earlier sections there are a minority of labels corresponding to benign and Web DDoS. Here the process of synthetic minority oversampling technique (SMOTE) [14] is used to rectify the minorities and make their distribution equal to the other classes. This process is carried out separately for both DDoS and 5G input features. The Figures 5 and 6 below shows the new distribution of the classes and the total number of values associated to them. It is evident that all the minorities are rectified.

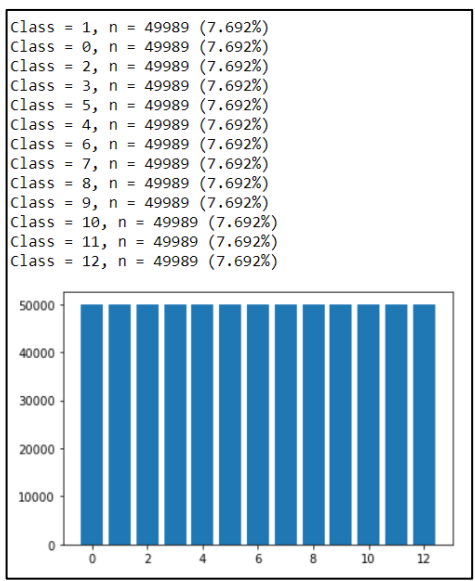

*Figure 4: Oversampling DDoS labels results*

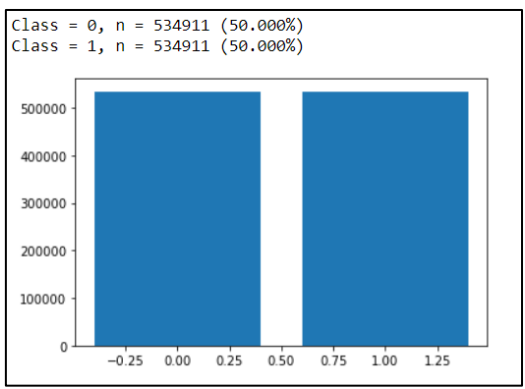

*Figure 4: Oversampling 5G latency label results*

### 4. Feature Selection

Machine learning is a delicate process. If the data is not preprocessed properly, it will affect the performance of the models. Since the dataset is very large with over 80 features, there is a high chance that feeding all the features as inputs will confuse the model and the predictions will be drastically affected. For this reason, a two-step feature selection process is considered.

#### 4.1. Feature selection – step 1: K Best Features

As stated, the first stage of feature selection employs the K best features approach. This method is adopted as a crude form of feature selection to get the 40 best features. This is method is first adopted because it is faster in processing the data and returning the output. K best features is class of the scikit learn feature selection library. The score function used is the f_regression method. It is a univariate linear regression test that return F-statistic and p-values. It ranks the features in the same order if they are positively correlated to the target.

#### 4.2. Feature selection – step 2: Recursive Feature Estimation (RFE)

The second stage of feature selection employs the Recursive Feature Estimation technique. This is used as the last step of feature selection to get the refined final 20 features that will directly be fed to the machine learning models. The RFE is also a part of the scikit learns feature estimation library. The recursive method is relatively time consuming and has a slight overhead. For this reason, it was adopted as the second stage to reduce the time taken to process the data. The estimator passed to the RFE was the Decision Tree Regressor. The final features were then stored in a final variable called X DDoS and 5G train/test and y DDoS and 5G train/test.

## RESULTS AND ANALYSIS

This section deciphers the performance of all the models after training was completed. First, a list is made consisting of all the names of the classifiers that were used in the modeling process. Next a list named scores_ddos is created which holds a list accuracy score, precision score, f1 score and recall score of all the classifiers used for DDoS predictions. Similarly, a list named scores_5G is created which holds information of scores of all the classifiers used for 5G predictions. The plots are visualized in the next section.

1. Model evaluation

The plots are created with the axes showing all the classifiers and how they performed. Figure 7 show the plot for all the classifiers for DDoS.

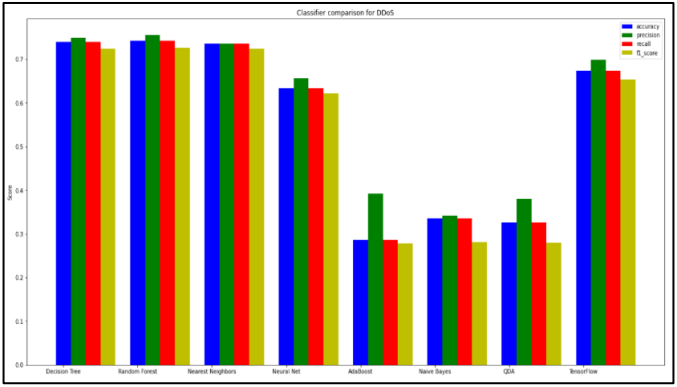

*Figure 7: DDoS attack classifier comparison*

From the above figure it is observed that the Random Forest classifier performs the best with the accuracy peaking at around 74 percent. The overall performance of all the other scores of the classifier are also relatively good. This means that adopting the random forest is a good choice in predicting the output of the type of DDoS attack. The Decision Tree, Random Forest and TensorFlow classifiers are not far behind. The models can be tweaked further to improve the accuracy and other scores of the models. Adding extra hidden layers in the deep neural net could be beneficial. In particular, the random forest estimators can be increased to get higher accuracy. Running recursive estimation could help identify even better features to feed to the model. Similarly, the figure 8 below shows the plot of all the 5G label classifiers and their performance.

From the above figure it is observed that the AdaBoost classifier performs the best with the accuracy peaking at around 53 percent. The overall performance of all the other scores of the classifier are also relatively good. This means that adopting the AdaBoost classifier gives the highest chance in predicting the output of the type of 5G Latency Label. The other classifiers are not far behind. The models can be tweaked further to improve the accuracy and other scores of the models. Adding extra hidden layers in the deep neural net or increasing the nearest neighbors could be beneficial. In particular, the AdaBoost estimators can be increased to get higher accuracy. Running recursive estimation could help identify even better features to feed to the model. Similarly, the figure 14 below shows the plot of all the classifiers and their performance.

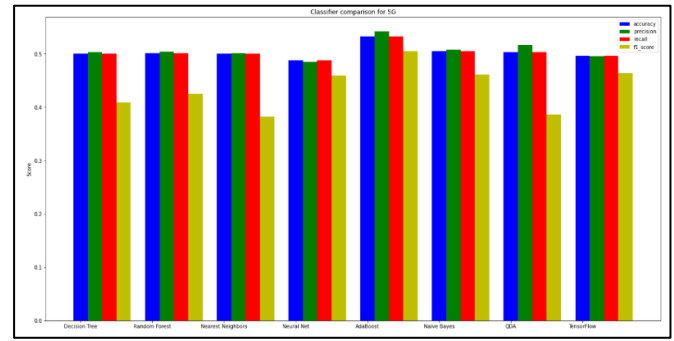

*Figure 8: 5G Latency label classifier comparison*

The classifier selection process is important as each classifier algorithm behavior is different. A lot of real-world testing must be conducted before any other decisions are made.

# CONCLUSION

The main goal this thesis implementation was to understand the behavior of 5G latency with respect to an attack. This implementation started out with exploring and analyzing the initial dataset and its contents. A visual analysis of the DDoS attack types and the 5G components is displayed and described in brief. Some of the important features are also explored and visualized. It is observed that the flow duration, ACK flag count, total forward packets, flow packets per second, total backward packets, protocol, source, and destination ports are some of the main features. After visualizing the features, they are split into inputs and outputs for machine learning. Since the 5G labels and benign data was not equally balanced, the observations were oversampled to match the other majority classes. The data was then scaled and normalized to make it easier for the machine learning models to interpret. Finally, the data was split into training and testing. A total of eight machine learning models were applied and the best ones were visualized. Therefore, a machine learning model to correctly predict the impact of an attack on the 5G latency of a vehicle was developed.

The research was based on largely quantitative research methods in a simulated environment. Hence, the biggest limitation of this research was the lack of resources for data collection and sole reliance on online data sets. Ideally a V2X project would greatly benefit from an autonomous 5G enabled vehicle connected to a mobile edge cloud. However, this project was conducted solely online on a single PC which further limits the outcomes.

5G and cloud service providers are working hand in hand to try and make a resilient 5G vehicular network. One of the main challenges they face is in securing these networks. The aim of this thesis project is helping the infrastructure providers secure the 5G V2X system by applying machine learning for attack detection and prediction of 5G latency. This can be a steppingstone to unlock the full potential of an intelligent transport system that can prevent road accidents.